\documentstyle[12pt,epsfig]{article} 
\newlength{\dinwidth}                       
\newlength{\dinmargin}                      
\def\lsim{\mathrel{\rlap{\lower4pt\hbox{\hskip1pt$\sim$}}
    \raise1pt\hbox{$<$}}}                
\def\gsim{\mathrel{\rlap{\lower4pt\hbox{\hskip1pt$\sim$}}
    \raise1pt\hbox{$>$}}}                
\newcommand\beq{\begin{equation}}
\newcommand\eeq{\end{equation}}
\newcommand\bea{\begin{eqnarray}}
\newcommand\eea{\end{eqnarray}}
\begin{document}

\begin{titlepage}

\begin{flushleft}
DESY 96--199 \\[0.1cm] 
INLO-PUB-20/96 \\[0.1cm]
WUE-ITP-96-018 \\[0.1cm] 
{\tt hep-ph/9609400} \\[0.1cm]
September 1996
\end{flushleft}
\vspace{0.2cm}
\begin{center}
\Large
{\bf A Detailed Comparison of NLO QCD Evolution Codes} \\
\vspace{1.2cm}
\large
J. Bl\"umlein, S. Riemersma \\
\vspace{0.3cm}
\normalsize
{\it DESY--Zeuthen, Platanenallee 6, D--15735 Zeuthen, Germany} \\
\vspace{0.8cm}
\large
M. Botje \\
\vspace{0.3cm}
\normalsize
{\it NIKHEF, PO Box 41882, 1009 DB Amsterdam, The Netherlands} \\
\vspace{0.8cm}
\large
C. Pascaud, F. Zomer \\
\vspace{0.3cm}
\normalsize
{\it LAL, IN2P3-CNRS et Universit\'e de Paris-Sud} \\
\vspace{0.1cm}
{\it F--91495 Orsay C\'edex, France} \\
\vspace{0.8cm}
\large
W.L. van Neerven \\
\vspace{0.3cm}
\normalsize
{\it Instituut Lorentz , Rijksuniverstiteit Leiden} \\
\vspace{0.1cm}
{\it PO Box 9506, 2300 RA Leiden, The Netherlands}\\
\vspace{0.8cm}
\large
A. Vogt\\
\vspace{0.3cm}
\normalsize
{\it Institut f\"ur Theoretische Physik, Universit\"at W\"urzburg} \\
\vspace{0.1cm}
{\it Am Hubland, D--97074 W\"urzburg, Germany} \\
\vspace{1.2cm}
\large
{\bf Abstract}
\normalsize
\end{center}
\vspace{-0.3cm}
Seven next-to-leading order QCD evolution programs are compared. The
deviations of the results due to different theoretical prescriptions
for truncating the perturbative series are clarified, and a numerical
agreement between five codes of better than 0.1\% is achieved.
Reference results for further comparison are provided.
\vfill 
\noindent
\small
{\it To appear in the proceedings of the workshop ``Future Physics at 
HERA'', DESY, Hamburg, 1996. }
\normalsize
 
\end{titlepage}
%
\vspace*{1cm}
\begin{center}  
  \begin{Large} \begin{bf}
A Detailed Comparison of NLO QCD Evolution Codes\\
  \end{bf}  \end{Large}
\vspace*{5mm}
  \begin{large}
J. Bl\"umlein$^a$, M. Botje$^b$, C. Pascaud$^c$, S. Riemersma$^a$, \\ 
W.L. van Neerven$^d$, A. Vogt$^{e,f}$, and F. Zomer$^c$
  \end{large}
\end{center}
$^a$ DESY-Zeuthen, Platanenallee 6, D--15738 Zeuthen, Germany\\
$^b$ NIKHEF, PO Box 41882, 1009 DB Amsterdam, The Netherlands\\
$^c$ LAL, IN2P3-CNRS et Universit\'e de Paris-Sud, F--91495 Orsay C\'edex, 
     France\\
$^d$ Instituut Lorentz, Rijksuniversiteit Leiden, PO Box 9506, 2300 RA
     Leiden, The Netherlands\\
$^e$ Deutsches Elektronen-Synchrotron DESY, Notkestra{\ss}e 85, 
     D--22603 Hamburg. Germany\\
$^f$ Institut f\"ur Theoretische Physik, Universit\"at W\"urzburg,
     D--97074 W\"urzburg, Germany

\begin{quotation}
\noindent
{\bf Abstract:}
Seven next-to-leading order QCD evolution programs are compared. The 
deviations of the results due to different theoretical prescriptions
for truncating the perturbative series are clarified, and a numerical 
agreement between five codes of better than 0.1\% is achieved. 
Reference results for further comparison are provided.
\end{quotation}
 
\section{Introduction}

In order to exploit the full potential of HERA for deep--inelastic
scattering (DIS), the highest possible luminosities and considerable 
efforts for the reduction of experimental systematic uncertainties are 
necessary. This will finally allow a measurement of the proton structure
function $F_2$ over a wide range, with errors on the level of very few 
percent \cite{BKP}. To make full use of such results, and to allow 
even for combined analyses using the high--precision fixed--target 
data as well, the structure function evolution programs required 
for the necessary multi--parameter QCD fits have to be numerically and 
conceptually under control to a much higher accuracy. At least one 
order of magnitude is desirable.
This accuracy is necessary to safely rule out contributions to the
theory error of $\alpha_s(M_Z^2)$ which arise from the particular
technical implementation of the solution of the NLO evolution
equations. Due to the current apparent difference in $\alpha_s(M_Z^2)$
as determined in $e^+e^-$ and DIS experiments~\cite{JBR}, this question 
is of particular importance for the future QCD analyses based on the 
HERA structure function data.
 
So far no high--precision comparison of next-to-leading-order (NLO)
programs has been performed including the full HERA range. 
In previous studies partial comparisons were carried out demanding
a considerably lower accuracy (see e.g.\ ref.~\cite{Tung}). Other 
comparisons focussed on the valence range and compared the effect 
of different codes used for the QCD fit on $\Lambda_{\rm QCD}$ only, 
see ref.~\cite{BCDMS}. The required accuracy cannot be easily reached 
by just comparing results to published parametrizations, due to their 
inaccuracies, caused by respective numerical representations. Often 
also the physical and technical assumptions made are not fully 
documented.

In this paper, we present the results of a dedicated effort, 
comparing the results of seven NLO codes under perfectly controlled 
conditions. The paper is organized as follows. In Section~2 we recall 
the basic formulae, and sketch the most commonly used approaches to 
the evolution equations. Section 3 compares the differences of six 
`global' evolution programs. The clarification of the deviations found 
there, using also a seventh program especially suited for the `local' 
evolution of $F_2$, is described in Section 4. The size of the 
numerical differences which persist after this development is 
investigated in Section 5, where also reference results for further 
comparison are provided. Finally Section 6 contains our summary.

\section{Approaches to the Next-to-Leading Order Evolution}

The evolution equations for the parton distributions $f(x,Q^2)$ of the 
proton are given by
\beq
\label{evol}
 \frac{\partial f(x,Q^2)}{\partial \ln Q^2 } = \left[ a_{s}(Q^2)\, 
 P_{0}(x) + a_s^2(Q^2)\, P_1(x) + O(a_s^3) \right] \otimes f(x,Q^2)\: .
\eeq
Here $x$ stands for the fractional momentum carried by the partons, 
and $\otimes $ denotes the Mellin convolution. For brevity, we have 
introduced $ a_s(Q^2) \equiv \alpha_s(Q^2) / 4\pi$. 
Eq.\ (\ref{evol}) is understood to represent, in a generic manner, the 
non--singlet cases as well as the coupled quark and gluon evolutions. 
$P^{(0)}$ and $P^{(1)}$ denote the corresponding leading order (LO) and 
NLO splitting functions, respectively (see, e.g., ref.\ \cite{WvNrev}). 
Only these two coefficients of the perturbative series are completely 
known so far, hence the solution of the evolution equations is 
presently possible only up to NLO.
To this accuracy, the scale dependence of the strong coupling $a_s(Q^2)$
reads 
\beq
\label{arun}
 \frac{\partial \, a_s(Q^2)}{\partial \ln Q^2 } =
 - \beta_0 a_s^2(Q^2) - \beta_1 a_s^3(Q^2) + O(a_s^4) \: .
\eeq
Throughout our comparisons, we will identify the renormalization and
factorization scales with $Q^2$, as already indicated in eqs.\ 
(\ref{evol}) and (\ref{arun}). For different choices see refs.\ 
\cite{BRNVrhe,BRNVhera}. Introducing the QCD scale parameter $\Lambda $,
the solution of eq.\ (\ref{arun}) can be written as
\begin{equation}
\label{arun1}
 a_s(Q^2) \simeq  \frac{1}{\beta_0 \ln(Q^2/\Lambda^2)}
 - \frac{\beta_1}{\beta_0^3} \frac{\ln\, [\ln(Q^2/\Lambda^2)]}
   {\ln^2 (Q^2/\Lambda^2)} \: .
\end{equation}

Two approaches have been widely used for dealing with the integro--%
differential equations (\ref{evol}). In many analyses, they have been 
numerically solved directly in $x$-space. We will exemplify some
techniques applicable in this case for one particular program, choosing
`{\sc Qcdnum}', which is based on the programs of ref.\ \cite{VO}, and 
is planned to become publicly available \cite{MB}. See, for example, 
ref.\ \cite{PZ} for a description of a differing $x$-space 
implementation.

In {\sc Qcdnum}, the $Q^2$ evolution of the parton momentum densities 
is calculated on a grid in $x$ and $Q^2$, starting from the 
$x$-dependence of these densities at a fixed reference scale $Q^2_0$.
The logarithmic slopes in $Q^2$ are calculated from eq.\ (\ref{evol}). 
To compute the convolution integrals, the assumption is made that the 
parton distributions can be linearly interpolated (at all $Q^2$) from 
one $x$ gridpoint to the next. With this assumption the integrals can 
be evaluated as weighted sums. The weights, which are essentially 
integrals over the splitting functions, are numerically calculated (by 
Gauss integration) to high precision at program initialization.
From the value of a given parton distribution and the slopes at $Q^2_0$,
the distribution can be calculated at the next gridpoint $Q^2_1 > Q^2_0$
(or $Q^2_1 < Q^2_0$). This distribution then serves to calculate the 
slopes at $Q^2_1$ etc., and the evolution is continued over the whole
$x$--$Q^2$ grid. The evolution algorithm makes use of quadratic 
interpolation in $\ln Q^2.$

In this way, a fast evolution of parton densities is obtained, entirely 
based on look-up weight tables which are calculated at program 
initialization. The numerical accuracy depends on the density of the 
$x$ grid and, to a lesser extent, on that of the $Q^2$ grid. In the 
comparisons presented here, 370 gridpoints in $x$ covering $10^{-5} 
\leq x < 1$ have been used: 230 points distributed logarithmically for 
$x < 0.2$, and 140 points distributed linearly for $x > 0.2.$ A 
logarithmic $Q^2$ grid with 60 points covered the range $4 < Q^2 < 
10^4$ GeV$^2$.

An important alternative to the direct $x$-space treatment, employed 
in the analyses of refs.\ \cite{DFLM,GRV90} based upon ref.\ \cite
{FP}, is to transform the evolution equations to Mellin-$N$ moments. 
The main virtue of this transformation is that the convolution is 
reduced to a simple product. Hence eq.\ (\ref{evol}) turns into a 
system of ordinary differential equations at fixed $N$, which allows
for an analytic solution. Rewriting the evolution equations in terms 
of $a_s \equiv a_s(Q^2)$ using eq.\ (\ref{arun}), and expanding the 
resulting r.h.s.\ into a power series in $\alpha_s$, one arrives in NLO
at
\begin{equation}
 \frac{\partial f(x,a_s)}{\partial a_s} = - \frac{1}{\beta_0 a_s}
 \bigg[ P^{(0)}(x) + a_s \bigg( P^{(1)}(x) - \frac{\beta_1}{\beta_0}
 P^{(0)}(x) \bigg) + O(a_s^2) \bigg] \otimes f(x,a_s) \: .
\label{evol1}
\end{equation}
After transformation to $N$-moments, its solution can be written down
in a closed form for the non--singlet cases, with $a_0 \equiv 
a_s(Q_0^2)$, as
\begin{equation}
 f_N(a_s) = \bigg [ 1 - \frac{a_s - a_0}{\beta_0} \bigg ( P^{(1)}_N - 
 \frac{\beta_1}{\beta_0} P^{(0)}_N \bigg ) + O(a_s^2) \bigg ]
 \bigg ( \frac{a_s}{a_0} \bigg )^{P^{(0)}_N/\beta_0} f_N(a_0) \: .
\label{sol}
\end{equation}
For the notationally more cumbersome, corresponding relation for the 
singlet evolution, the reader is referred to refs.\ \cite{DFLM,GRV90}. 

From these analytic solutions, one can acquire the $x$-space results by 
one contour integral in the complex $N$-plane, see ref.\ \cite{GRV90}. 
Using a chain of Gauss quadratures, a numerical accuracy of this 
integration at better than $10^{-5}$ is readily achieved. In our 
comparisons, at most 136 fixed support points at complex $N$-values 
have been used, with this maximal number employed only for very large 
values of $x$ \cite{AV}.  
Due to the required non--trivial analytic continuations of the NLO
anomalous dimensions \cite{GRV90}, this approach is technically 
somewhat more involved than the numerical $x$-space solution. On the 
other hand, since the $Q^2$ integration is done in one step, regardless
of the evolution distance, and the use of fixed support points allows 
for performing the calculation of the anomalous dimensions only once 
at program initialization, this method is competitive in speed to 
the $x$-space iterations.

A partly independent $N$-space program has been developed during this 
workshop \cite{SR}, implementing an iterative numerical solution of 
the Mellin--transformed eq.\ (\ref{evol1}). Since one of the advantages 
of the $N$-space approach is not exploited here, this program is so 
far not competitive in speed with the ones discussed before. It has 
however been of considerable value for cross--checks and theoretical 
investigations, see below.

Before we now turn to the comparisons, it should be emphasized that 
a perfect agreement between the results based upon eqs.\ (\ref{evol}), 
(\ref{evol1}), and (\ref{sol}) is not to be expected, since they all 
differ in terms of next-to-next-to leading order (NNLO), hidden under 
the $O(a_s^3)$ and $O(a_s^2)$ signs. 

\section{The Initial Comparisons}

All our comparisons are performed under somewhat simplified, but 
sufficiently realistic conditions. We assume four massless flavours,
in eq.\ (\ref{evol}) as well as in eq.\ (\ref{arun}), at all scales
considered, i.e.\ effects due the non--zero charm mass and the existence
of the bottom quark are not taken into account. All our results below
will refer to the $\overline{\rm MS}$ renormalization and factorization
schemes, and the corresponding scales are identified with $Q^2$. The 
reference scale $Q_0^2$ for the evolution, and the four--flavour QCD 
scale parameter $\Lambda $ in eq.\ (\ref{arun1}) are chosen as
\beq
\label{lam}
  Q_0^2 = 4 \mbox{ GeV}^2 \: , \:\:\: 
  \Lambda_{\overline{\rm MS}}^{(4)} = 250 \mbox{ MeV} \: .
\eeq
\begin{figure}[b]
\vspace*{-5mm}
\begin{center}
\mbox{\epsfig{file=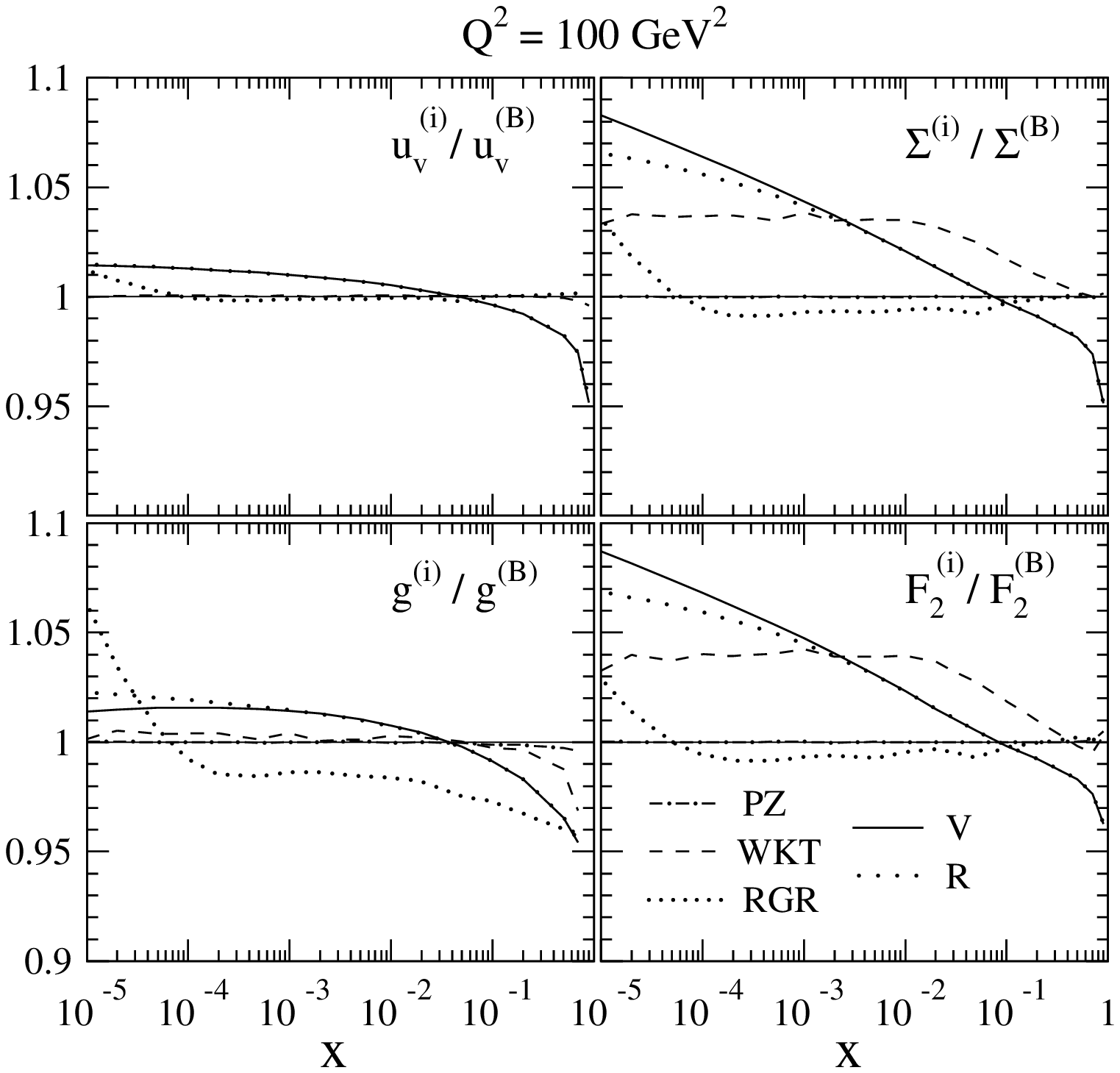,height=13.2cm}}
\end{center}
\vspace*{-6mm}
{\sf Figure~1:~The differences between the up-valence, singlet quark
 and gluon densities, $u_v$, $\Sigma $ and $g$, and the proton structure
 functions $F_2$, as obtained from evolving the input (\ref{inp}) with 
 various NLO evolution programs \cite{MB,PZ,AV,SR,RGR,WKT} to $Q^2 = 
 100$ GeV$^2$. All results have been normalized to those of ref.\ \cite
 {MB}. For a detailed discussion see the text.}
\end{figure}
The following initial conditions are selected for the (anti-) 
quark and gluon densities:  
\bea
\label{inp}
 xu_v(x,Q_0^2) \! &\! =\! & A_u x^{0.5} (1-x)^3 \: , \:\: 
 xd_v(x,Q_0^2) \, = \, A_d x^{0.5} (1-x)^4 \: , \nonumber\\
 xS(x,Q_0^2) &\! =\! & [x\Sigma \, - \, xu_v \, - \, xd_v](x,Q_0^2) \:
  \:\:\: = \, A_S x^{-0.2} (1-x)^7 \: , \\
 xg(x,Q_0^2) &\! =\! & A_g x^{-0.2} (1-x)^5 \: , \:\: 
 xc(x,Q_0^2) \, = \, x\bar{c}(x,Q_0^2) \: = \: 0 \: .  \nonumber
\eea
The SU(3)--symmetric sea $S$ is assumed to carry 15\% of the nucleon's 
momentum at the input scale, and the remaining coefficients $A_i$ are 
fixed by the usual sum rules. Finally $F_2$ is determined by simply 
convoluting the resulting parton densities with the appropriate 
coefficient functions.

The results of our first comparisons are shown in Figure 1. One notices
the very good agreement between the programs \cite{MB,PZ} used by the 
HERA collaborations. The differences are always much less than 1\%, and 
the curves can hardly be distinguished, except for large $x$, with the 
present resolution. A similarly excellent agreement is seen between the 
two $N$-space programs \cite{AV,SR}, except for very low $x$, where
offsets up to 1.5\% show up. The most striking feature of the figure,
however, is the very sizeable differences between these two groups of
programs: the scaling violations, increasing (decreasing) the 
distributions at small (large) values of $x$, are considerably stronger
in the results of refs.\ \cite{AV,SR}, although, of course, the same 
values for $\alpha_s$ are employed as in refs.\ \cite{MB,PZ}.
This effect reaches a magnitude of as much as 8\% for the structure 
function $F_2$ at the smallest $x$-values considered. 

As stated above, perfect agreement had not been expected due to 
theoretical differences, but the size of this offset was a surprise to 
most of us. It initiated quite some checking and programming activity, 
which will be summarized in the next section.

Also shown in the figure are the results obtained by the $x$-space 
evolution programs of the MRS and CTEQ global fit collaborations 
\cite{RGR,WKT}. Very good agreement to the results of refs.\ \cite
{MB,PZ} is found for the valence quarks, except for ref.\ \cite{RGR} 
at extremely low values of $x$. In the singlet sector, however, 
significant differences are observed for some quantities: 1.5 -- 3\% on 
the gluon density in ref.\ \cite{RGR}, and up to 4\% on the sea quark 
distributions in ref.\ \cite{WKT}.

\section{Pinning Down the Differences}

Besides checks and comparisons of the numerical values of the NLO
splitting between the codes of refs.\ \cite{MB,AV,SR}, the large 
differences discussed in the previous section led to three program
developments, which together allowed for their full understanding
on an unprecedented level.
\begin{itemize}
\item A program for a local representation of the evolution of 
 $F_2$ close to the initial scale, completely independent of all 
 previous ones, was added to the comparisons \cite{BvN}.
\item The code of ref.\ \cite{SR} was extended to include, still
 in moment space, an option for evolving also on the basis of eq.\ 
 (\ref{evol}) instead of (\ref{evol1}).
\item The program of ref.\ \cite{AV} was used to simulate an 
 iterative solution of eq.\ (\ref{evol1}) as performed in ref.\ 
 \cite{SR}, and additionally two new iterative options, one of them 
 equivalent to eq.\ (\ref{sol}), were introduced into this package.
\end{itemize}

The results of these efforts are displayed in Figure 3, where we show
the evolution of $F_2$, close to our reference scale $Q_0^2 = 4 $ 
GeV$^2$, for three typical values of $x$. The differences, depicted in 
the previous figure for $Q^2 = 100 $ GeV$^2$, build up very quickly 
near $Q_0^2$: already around 10 GeV$^2$ they are close to their final 
level. 
The second important observation is the perfect agreement of the local 
representation \cite{BvN} with the $x$-space codes \cite{MB,PZ}, which 
immediately stopped any speculations on possible problems in the latter 
programs. Next one notices that the 1\% small-$x$ difference between 
refs.\ \cite{AV} and \cite{SR} is perfectly understood in terms of the 
slightly different contributions truncated away in eqs.\ (\ref{evol1}) 
and (\ref{sol}), cf.\ ref.\ \cite{BRNVrhe}. 
The concluding step is the comparison of the modified evolution of 
\cite{SR} with the results of \cite{MB,PZ,BvN}. This reveals that in 
fact virtually all offsets between the results of refs.\ 
\cite{MB,PZ,AV,SR} in Figure 1 are due to the differences introduces 
by the employed truncation prescriptions for the perturbative series 
the NLO level, i.e.\ by terms of NNLO and beyond.

\begin{figure}[tbh]
\vspace*{-5mm}
\begin{center}
\mbox{$\!$\epsfig{file=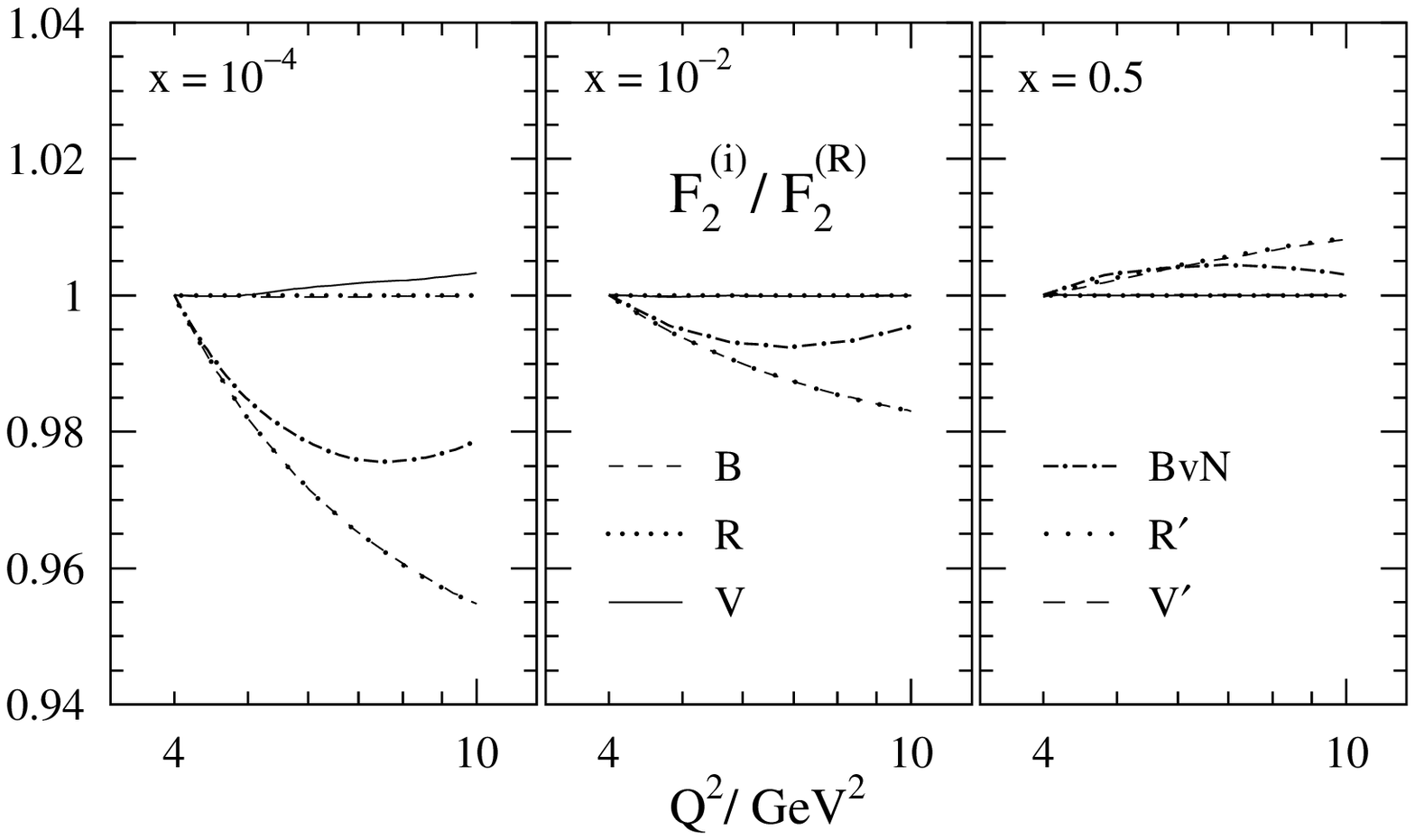,height=10cm}$\!$}
\end{center}
\vspace*{-4mm}
{\sf Figure~2:~A comparison of results on the $Q^2$ evolution of $F_2$
 close to the reference scale $Q_0^2 = 4 $~GeV$^2$. The results of 
 refs.\ \cite{MB,SR,GRV90} (denoted by {\bf B}, {\bf R}, and {\bf V}) 
 are as in the previous figure. {\bf BvN} represents a local 
 representation of the $F_2$ evolution \cite{BvN}, and the curves {\bf 
 R$^{\prime}$} and {\bf V$^{\prime}$} check the numerical consistency 
 by adapting to the theoretical assumptions of refs.\ \cite{MB} and 
 \cite{SR}, respectively.}
\end{figure}

The origin of the differences between the results of refs.\ 
\cite{RGR,WKT} and our programs could not be clarified during this 
workshop. Hence for the very precise comparisons to which we now turn, 
we will keep only our five program packages, which agree, at least, 
sizeably better than to 1\%.

\section{The Achieved Numerical Accuracy}

Armed now with at least two different codes for any of the truncation
prescriptions of Section~2, we can proceed to explore the limits of 
the agreement of our five program packages under consideration. This 
complete coverage will be used for comparing all programs, even those 
with conflicting theoretical treatments, in one figure. For this 
purpose, the results of refs.\ \cite{MB,PZ} have been normalized to the
modified evolution of ref.\ \cite{SR} (based upon eq.\ (\ref{evol})), 
whereas the `iterated' evolution of ref.\ \cite{AV} is normalized to 
the original results of ref.\ \cite{SR} (based upon eq.~(\ref{evol1})).

The results are shown at two fixed $Q^2$ values in Figure 3 for the 
parton distributions, and in Figure 4 for $Q^2$ evolution of the proton 
structure function $F_2$ at three fixed values of $x$. The total spread
of the results at $Q^2 = 100$ GeV$^2$ amounts to at most about 0.05\%, 
except for very large $x$, where the distributions, especially the gluon
density, become very small. Even after evolution to $10^4$ GeV$^2$, the
differences are still on the level of 0.1\%, meeting the goal formulated
in the introduction. Moreover, there is no reason for failing to reach 
an even higher accuracy, at least to 0.02\% as already achieved between 
the $N$-space programs, also in $x$-space, e.g.\ by increasing the 
still not too high number of $Q^2$ grid points in the program of ref.\ 
\cite{MB}.

Finally, for the convenience of those readers who want to check their 
own existing of forthcoming NLO evolution program to an accuracy well 
below 0.1\% over a wide range in $x$, we show in Table 1 two sets of 
reference results, which represent the evolution of the initial 
distributions (\ref{inp}) under the conditions (\ref{lam}), according 
to eq.\ (\ref{evol}) and eq.\ (\ref{sol}) to $Q^2 = 100$ GeV$^2$.

\begin{figure}[thbp]
\begin{center}
\mbox{\epsfig{file=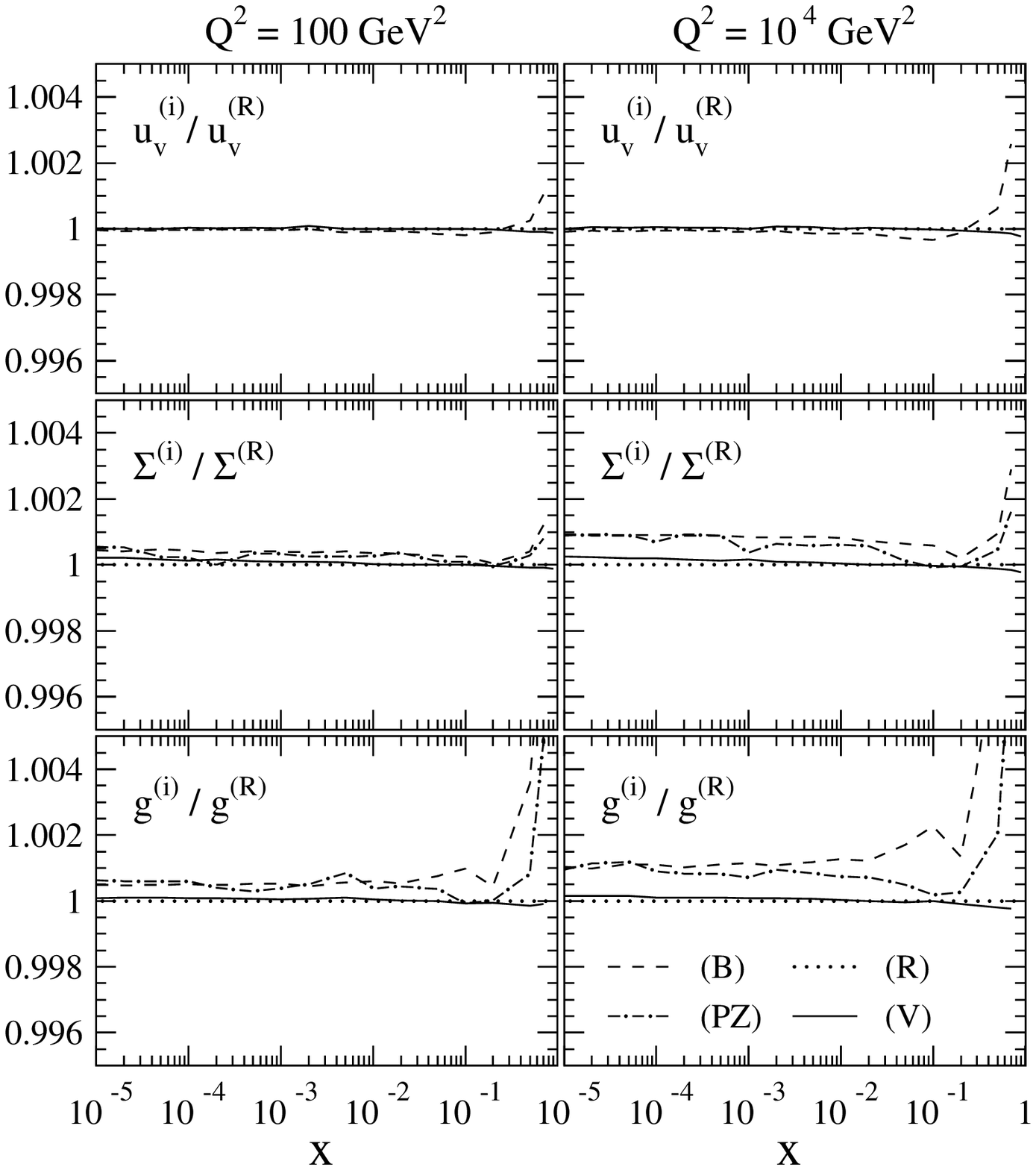,height=18.2cm}}
\end{center}
\vspace*{-6mm}
{\sf Figure~3:~The remaining numerical $x$-dependent deviations on 
 $u_v$, $\Sigma $ and $g$ at $Q^2 = 100$ and $10^4$ GeV$^2$ between the
 programs of refs.\ \cite{MB,PZ,AV,SR}, after the differing theoretical 
 assumptions have been corrected for, see the text. The results have 
 been normalized to those of ref.\ \cite{SR}.}
\end{figure}

\begin{figure}[thbp]
\begin{center}
\mbox{\epsfig{file=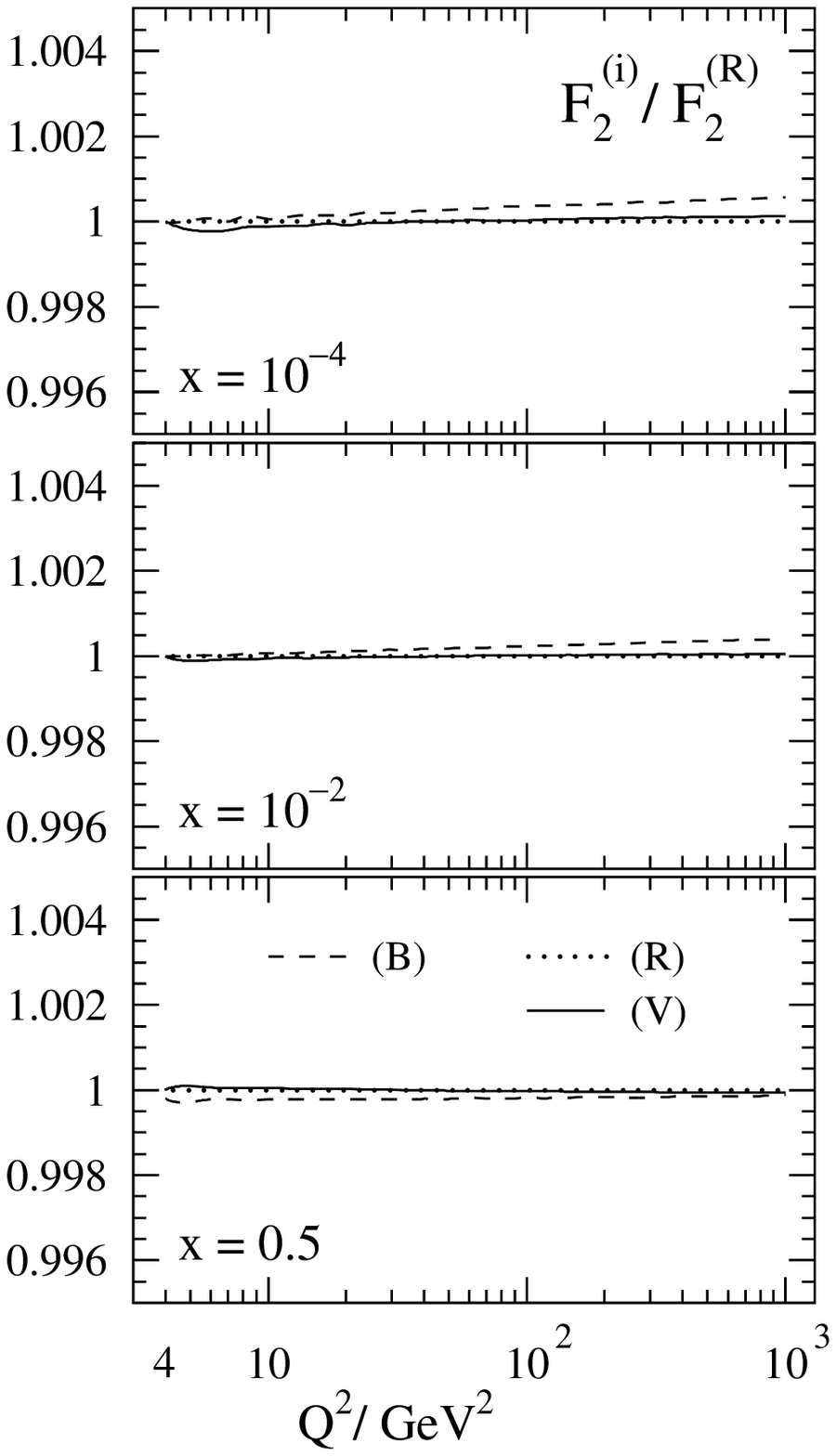,height=18.2cm}}
\end{center}
\vspace*{-4mm}
{\sf Figure~4:~The residual relative offsets in the $Q^2$ evolution of 
 $F_2$ between the programs of refs.\ \cite{MB,AV,SR}, after removing 
 the effects due to the different theoretical NLO prescriptions, for 
 three typical values of $x$. As in all previous figures, the initial 
 distributions are taken from eq.\ (\ref{inp}).} 
\end{figure}

\begin{center}
\vspace{2mm}
\begin{tabular}{||c||l|l|l|l|l|l||}
\hline \hline
 & & & & & & \\[-0.3cm]
\multicolumn{1}{||l||}{$x$} &
\multicolumn{1}{c|} {$xu_v$} &
\multicolumn{1}{c|} {$xd_v$} &
\multicolumn{1}{c|} {$xS$} &
\multicolumn{1}{c|} {$2xc$} &
\multicolumn{1}{c|} {$xg$} &
\multicolumn{1}{c||}{$F_2^{\, }$} \\
 & & & & & & \\[-0.3cm] \hline \hline
 & & & & & & \\[-0.3cm]
 $10^{-5}$ & 9.2793$\,$E-3 & 5.2115$\,$E-3 & 2.6670$\,$E1 
	   & 5.0866$\,$E0  & 9.6665$\,$E1  & 7.0270$\,$E0 \\
 $10^{-4}$ & 2.8777$\,$E-2 & 1.6134$\,$E-2 & 1.3862$\,$E1 
	   & 2.4694$\,$E0  & 4.7091$\,$E1  & 3.5868$\,$E0 \\
 $10^{-3}$ & 8.7208$\,$E-2 & 4.8678$\,$E-2 & 6.7508$\,$E0 
	   & 1.0663$\,$E0  & 2.0801$\,$E1  & 1.7271$\,$E0 \\
 $10^{-2}$ & 2.4598$\,$E-1 & 1.3494$\,$E-1 & 2.8562$\,$E0 
	   & 3.5762$\,$E-1 & 7.5998$\,$E0  & 7.9497$\,$E-1 \\
 0.1       & 4.7450$\,$E-1 & 2.3215$\,$E-1 & 5.7924$\,$E-1 
	   & 4.6496$\,$E-2 & 1.4260$\,$E0  & 3.5397$\,$E-1 \\
 0.3       & 3.1152$\,$E-1 & 1.1662$\,$E-1 & 5.7780$\,$E-2 
	   & 3.5268$\,$E-3 & 1.9173$\,$E-1 & 1.6536$\,$E-1 \\
 0.7       & 2.5048$\,$E-2 & 3.9486$\,$E-3 & 8.0219$\,$E-5 
	   & 4.0111$\,$E-6 & 1.1276$\,$E-3 & 1.4359$\,$E-2 \\
\hline \hline
& & & & & & \\[-0.3cm]
\multicolumn{1}{||c||}{$x$} &
\multicolumn{1}{c|} {$xu_v$} &
\multicolumn{1}{c|} {$xd_v$} &
\multicolumn{1}{c|} {$xS$} &
\multicolumn{1}{c|} {$2xc$} &
\multicolumn{1}{c|} {$xg$} &
\multicolumn{1}{c||}{$F_2^{\, }$} \\
& & & & & & \\[-0.3cm] \hline \hline
& & & & & & \\[-0.3cm]
 $10^{-5}$ & 9.4109$\,$E-3 & 5.2848$\,$E-3 & 2.8893$\,$E1 
	   & 5.6465$\,$E0  & 9.8060$\,$E1  & 7.6417$\,$E0 \\
 $10^{-4}$ & 2.9144$\,$E-2 & 1.6336$\,$E-2 & 1.4755$\,$E1 
	   & 2.6954$\,$E0  & 4.7859$\,$E1  & 3.8325$\,$E0 \\
 $10^{-3}$ & 8.8083$\,$E-2 & 4.9146$\,$E-2 & 7.0516$\,$E0 
	   & 1.1434$\,$E0  & 2.1110$\,$E1  & 1.8094$\,$E0 \\
 $10^{-2}$ & 2.4723$\,$E-1 & 1.3553$\,$E-1 & 2.9226$\,$E0 
	   & 3.7584$\,$E-1 & 7.6627$\,$E0  & 8.1358$\,$E-1 \\
 0.1       & 4.7268$\,$E-1 & 2.3097$\,$E-1 & 5.7880$\,$E-1 
	   & 4.7422$\,$E-2 & 1.4152$\,$E0  & 3.5337$\,$E-1 \\
 0.3       & 3.0798$\,$E-1 & 1.1511$\,$E-1 & 5.6817$\,$E-2 
	   & 3.4796$\,$E-3 & 1.8757$\,$E-1 & 1.6349$\,$E-1 \\
 0.7       & 2.4433$\,$E-2 & 3.8429$\,$E-3 & 7.6136$\,$E-5 
	   & 3.5004$\,$E-6 & 1.0854$\,$E-3 & 1.4013$\,$E-2 \\
\hline \hline

\end{tabular}
\end{center}
\noindent
{\sf Table 1. Reference results at $Q^2 = 100$ GeV$^2$ for the NLO 
 evolution using the direct solution of eq.\ (1) (upper half), and the 
 truncated analytic solution (5) (lower half). The initial conditions 
 are specified in eqs.\ (6) and (7). The estimated numerical accuracy 
 of these results is about 0.02\%.} 
\vspace{2mm}

\section{Summary}

The results of seven programs for the NLO evolution of parton densities
and structure functions have been compared. Differences due to terms
of NNLO, truncated differently in the various implementations, turn
out to be larger than anticipated. They can reach, e.g., about 6\% at 
$x = 10^{-4}$ and $Q^2 = 100$ GeV$^2$. A full quantitative understanding
of these differences has been achieved at an unprecedented level of
accuracy for five of these codes. There the remaining numerical 
differences are on the level of $\pm $0.02\% at  $Q^2 = 100$ GeV$^2$. 
Two sets of reference results, according to different theoretical 
prescriptions, have been provided for further high-precision checks of 
evolution programs.

\vspace{6mm}
\noindent
\large
{\bf Acknowledgements~:} \\[0.2cm]
\normalsize
We thank  R.G. Roberts  and  W.K. Tung  for providing the numerical
results of the evolution of our test input with their programs. 
This work was supported on part by the EC Network `Human Capital and 
Mobility' under contract No.\ CHRX--CT923--0004 and by the German 
Federal Ministry for Research and Technology (BMBF) under No.\ 05 
7WZ91P (0).


\begin{thebibliography}{99}
%
\bibitem{BKP}
M. Botje, M. Klein, and C. Pascaud, these proceedings.
%
\bibitem{JBR}
see e.g. J. Bl\"umlein, in: Proc. of the XXV International Symposium on
Multiparticle Dynamics, Stara Lesna, Slovakia, 11--17 September, 1995,
eds. D.~Bruncko, L.~Sandor, and J.~Urban,
(World  Scientific, Singapore, 1996), {\tt hep-ph/9512272}.
%
\bibitem{Tung}
W.K. Tung, Nucl.\ Phys.\ {\bf B315} (1989) 378.
%
\bibitem{BCDMS}
A.C. Benvenuti et al., BCDMS collaboration, Phys.\ Lett.\ {\bf B195} 
(1987) 97.
%
\bibitem{WvNrev}
W.L. van Neerven, these proceedings, and references therein.
%
\bibitem{BRNVrhe}
J. Bl\"umlein, S. Riemersma, W.L. van Neerven, and A. Vogt, DESY 
96--172, {\tt hep-ph/9609217}, Proceedings of the Workshop `QCD and QED 
in Higher Orders', Rheinsberg, Germany, April 1996, eds.\ J. Bl\"umlein,
F. Jegerlehner, and T. Riemann (Nucl. Phys. {\bf B} (Proc. Suppl.) 
{\bf 51C}, 1996) p.~96.
%
\bibitem{BRNVhera}
J. Bl\"umlein, S. Riemersma, W.L. van Neerven, and A. Vogt, these
proceedings.
%
\bibitem{VO}
M. Virchaux and A. Ouraou, DPhPE 87--15;\\
M. Virchaux, Th\`ese, Universit\'e Paris-7 (1988);\\
A. Ourau, Th\`ese, Universit\'e Paris-11 (1988).
%
\bibitem{MB}
M. Botje, {\sf QCDNUM15:  A fast QCD evolution program},
write-up in preparation.
%
\bibitem{PZ}
C. Pascaud and F. Zomer, H1 Note {\tt H1-11/94-404}.
%
\bibitem{DFLM} 
M. Diemoz, F. Ferroni, E. Longo, and G. Martinelli , Z.\ Phys.\ {\bf
C39} (1988) 21.
%
\bibitem{GRV90} 
M. Gl\"uck, E. Reya, and A. Vogt, Z. Phys.\ {\bf C48} (1990) 471.
%
\bibitem{FP} 
W. Furmanski and R. Petronzio, Z. Phys.\ {\bf C11} (1982) 293.
%
\bibitem{AV}
A. Vogt, unpublished.
%
\bibitem{SR}
S. Riemersma, unpublished.
%
\bibitem{RGR}
R.G. Roberts, contributed to this workshop.
%
\bibitem{WKT}
W.K. Tung, contributed to this workshop.
%
\bibitem{BvN}
J. Bl\"umlein and W.L. van Neerven, unpublished. For a short 
description, see Section 2.4 of ref.\ \cite{BRNVrhe}.
%
\end{thebibliography}
\end{document}